\documentclass[aps,prl,amssymb,showpacs,floatfix,twocolumn,reprint,nofootinbib]{revtex4-1}
\usepackage{amsmath,amsfonts,graphics,epsfig,amssymb,color}

\begin{document}
\title{Evolution of the Universe caused by the averaged potential of the quantum scalar field}

\author{Vladimir V. Belokurov}

\email{vvbelokurov@yandex.ru}

\affiliation{Lomonosov Moscow State University, Leninskie Gory 1, Moscow, 119991, Russia and Institute for Nuclear
Research of the Russian Academy of Sciences, 60th October Anniversary
Prospect 7a, Moscow, 117312, Russia}

\author{Evgeniy T. Shavgulidze}

\email{shavgulidze@bk.ru}

\affiliation{Lomonosov Moscow State University, Leninskie Gory 1, Moscow, 119991, Russia}

%\pacs{}

%\preprint{}
%\date{}

\begin{center}
\begin{abstract}
 We study the quantum theory of the singular scalar field $\varphi $ minimally coupled to gravity. In our approach, the scalar field is treated as a true quantum variable, while the scale factor $a(t)$ is supposed to be classical. We evaluate the quantum average of the self-interaction potential $V(\varphi)$ and use it in the modified differential equation for the Hubble parameter.
\end{abstract}
\end{center}
\maketitle

\vspace{1cm}

\section{1 Introduction}
\label{sec:intro}

The classical self-interacting scalar field minimally coupled to gravity is the key element of various models of the evolution of the Universe. (The papers are far too numerous to list here. The references can be found in many reviews, e.g., \cite{L}, \cite{Guth}, \cite{Linde}, \cite{Holten}.) The general action of the system is
  \begin{equation}
\label{Eq:Act}
S=\int\,\sqrt{-g}\left( -\frac{1}{2}R+\frac{1}{2}g^{\mu\nu} \partial_{\mu}\varphi \partial_{\nu}\varphi-V(\varphi) +\Lambda\right)d^{4}x\,.
\end{equation}
The standard cosmological framework is based on the FLRW metric $ds^{2}=dt^{2}-a^{2}(t)\,d\vec{x}^{2}\,,$ with flat three-dimensional space
where $t$ is the cosmic time and $a(t)$ is the scale factor.

The scalar field is assumed to be spatially constant: $\varphi(x^{\mu})=\varphi(t)\,.$
Under the above assumption, $\varphi(t) $  is not a true field but a mechanical variable. Nevertheless, we keep the name "field" for it.

For convenience, in this paper, we consider the Euclidean version of the theory \cite{CDL}, \cite{DH} given by the action
  \begin{equation}
\label{Eq:EAct}
S_{E}=\int\,\sqrt{g}\left( -\frac{1}{2}R+\frac{1}{2}g^{\mu\nu} \partial_{\mu}\varphi \partial_{\nu}\varphi+V(\varphi) -\Lambda\right)d^{4}x_{E}\,.
\end{equation}

   The most popular approach to studying the time evolution of the scale factor  is to use the equation of motion for the classical scalar field
\begin{equation}
\label{Eq:EM}
\ddot{\varphi}+3H\dot{\varphi}=-V'(\varphi)\,,
\end{equation}
and the Einstein equations
\begin{equation}
\label{Eq:Ein1}
3H^{2}=\frac{1}{2}\dot{\varphi}^{2}\,-\,V(\varphi)+\Lambda\,,
\end{equation}
\begin{equation}
\label{Eq:Ein2}
2\dot{H}+3H^{2}=-\frac{1}{2}\dot{\varphi}^{2}\,-\,V(\varphi)+\Lambda\,,
\end{equation}
simultaneously.
Here, $H$ is the Hubble parameter
$$
H=\frac{\dot{a}(t)}{a(t)}\,.
$$

It is well known that the equations  (\ref{Eq:EM}), (\ref{Eq:Ein1}), (\ref{Eq:Ein2}) are  redundant. In particular, for  $\dot{\varphi}\neq 0$ the last two equations imply the first one.

  In this paper, we suggest treating the scalar field as a true quantum variable, while the scale factor $a(t)$ is supposed to be classical. Now the equation of motion (\ref{Eq:EM}) is not valid for the quantum field $\varphi(t)$. Moreover, as we will see in the next section,
  the derivative $\dot{\varphi}(t)$ should be understood in a generalized sense, and the second derivative of the quantum field $\ddot{\varphi}(t)$ is not defined at all.
  Therefore, the equation (\ref{Eq:EM}) cannot be derived from the equations (\ref{Eq:Ein1}), (\ref{Eq:Ein2}).

The main idea of this paper is to modify the equation
\begin{equation}
\label{Eq:Ein12}
\dot{H}+3H^{2}=- V(\varphi(t))+\Lambda\,,
\end{equation}
that follows from (\ref{Eq:Ein1}), (\ref{Eq:Ein2}), for the case of the quantum scalar field.

  Namely, instead of dealing with the classical scalar field potential $V(\varphi(t))\,,$ we first evaluate its quantum average $\langle V(t)\rangle $  and then use it in the Riccati equation for $H$
\begin{equation}
\label{Eq:Riccati}
\dot{H}+3H^{2}=-\langle V(t)\rangle+\Lambda\,.
\end{equation}

The rest of the paper is organized as follows.
First, in Sec. $\bf{ 2}\,, $ we specify the model and reduce the problem of quantum averaging over the singular fields to the evaluation of Wiener integrals.
In Sec. $\bf{3}\,,$ we evaluate the Euclidean quantum average of the potential $V\,.$ Sec. $\bf{ 4}$ addresses the asymptotic solutions of the Riccati equation for $H\,.$ In Sec. $\bf{ 5}\,,$ we give the concluding comments.

\section{2 The model}
\label{sec:model}

One of the most popular inflationary potentials is the one arising from the Starobinsky $R^{2}$ model \cite{Star}, \cite{Barrow}:
\begin{equation}
\label{Eq:Starob}
V(\varphi)=V_{0}\left(1-e^{\alpha\varphi} \right)^{2}\,.
\end{equation}
Note that exponential scalar field potentials appear in many other approaches (see, e.g., \cite{Primordial}, \cite{Russo} and references therein).

We define the quantum averaged potential $ \langle V(t)\rangle $ as the functional integral
$$
\langle V(t)\rangle=N^{-1}\int\limits _{\varphi(0)=\varphi_{0}}V\left(\varphi(t) \right)
$$
\begin{equation}
\label{Eq:Q-average}
\times\,\exp\left\{-\int \limits _{0}^{+\infty}\left[\frac{1}{2}\dot{\varphi}^{2}+V(\varphi) \right] dt   \right\}\,d\varphi\,,
\end{equation}
with
$$
N=\int\limits _{\varphi(0)=\varphi_{0}}\exp\left\{-\int \limits _{0}^{+\infty}\left[\frac{1}{2}\dot{\varphi}^{2}+V(\varphi) \right] dt   \right\}\,d\varphi\,.
$$
 In what follows, we will not write the normalizing factor $N$ down explicitly.

The constant term $V_{0} $ in the exponent is cancelled by the same term in the normalizing factor $N^{-1}.$ Also, the presence of the constant $V_{0} $ in the pre-exponential factor in the integrand is eliminated by the shift in the cosmological constant. Therefore, we consider the equivalent potential of the form
\begin{equation}
\label{Eq:Potential}
V(\varphi)=-\frac{\alpha\lambda}{2}\,e^{\alpha\varphi}+\frac{\lambda^{2}}{2}\,e^{2\alpha\varphi}\,,\ \ \ \alpha<0, \, \lambda<0\,.
\end{equation}

We assume that there is an initial singularity and it is given by the initial condition $\varphi(0)=-\infty \,.$ As it follows from (\ref{Eq:Potential}),
$V(\varphi(0))=+\infty\,.$

Therefore, in the equation (\ref{Eq:Q-average}), the functional integration is performed over the space of functions continuous on the axis $(0,\,+\infty) $ except the initial point $t=0\,. $

Thus,  we have
$$
\langle V(t)\rangle=\int\limits _{\small{\begin {matrix} C((0,+\infty)) \\ \varphi(0)=-\infty \end {matrix}}} V\left(\varphi(t) \right)
$$
\begin{equation}
\label{Eq:E-average}
\times\exp\left\{-\int \limits _{0}^{+\infty}\left[\frac{1}{2}\dot{\varphi}^{2}+V(\varphi) \right] dt   \right\}\,d\varphi\,.
\end{equation}

Using the nonlinear nonlocal substitution
\begin{equation}
   \label{Eq:Subst}
\xi(t)=\varphi(t) - \varphi (0)+ \lambda\,\int \limits _{0}^{t}e^{\alpha \varphi (\tau)}d\tau\,,
\end{equation}
and its inverse
\begin{equation}
   \label{Eq:Inverse}
\varphi(t)=\xi(t)-\frac{1}{\alpha}\ln\left(\alpha \lambda \int \limits _{0}^{t}e^{\alpha \xi (\tau)}d\tau \right)
\end{equation}
we get the following equality of the functional integrals \cite{BSh} (see, also, \cite{BSh_arX}):
$$
\int\limits _{\small{\begin {matrix} C((0,+\infty)) \\ \varphi(0)=-\infty \end {matrix}}}V(\varphi(t))
$$
$$
\times \exp\left\{ - \frac{1}{2}\int \limits _{0}^{+\infty}\left \{ (\dot{\varphi}(t))^{2} + \lambda ^{2}\, e^{2\alpha \varphi (t)}- \alpha \lambda \, e^{\alpha \varphi (t)}\right \}dt \right\}\, d\varphi
$$
\begin{equation}
   \label{Eq:Equality}
= \int\limits _{\small{\begin {matrix} C([0,+\infty)) \\ \xi(0)=0 \end {matrix}}}V(\xi(t))\, w_{1}(d\xi) \,,
\end{equation}
where $ V(\xi(t))=V(\varphi(\xi(t)))\,, $ and the Wiener measure is written as
$$
w_{\sigma}(d\xi)=\exp\left\{ - \frac{1}{2\sigma^{2}}\int \limits _{0}^{+\infty} \dot{\xi}^{2}(t)\,dt \right\}\,   d\xi\,.
$$
As it is well known, the Wiener process $ \xi(t)$ is nondifferentiable. And in the above equation, the derivative  $\dot{\xi}(t) $ is understood in a generalized sense.

\section{3 The evaluation of Euclidean quantum average of the potential}
\label{sec:EQA}

Now we rewrite the potential $V$ in terms of $\xi \,.$
\begin{equation}
   \label{Eq:V}
V((\xi))=-\frac{1}{2} \frac{e^{\alpha\xi(t)}}{\int \limits _{0}^{t} \,e^{\alpha\xi(\tau)}d\tau }+\frac{1}{2\alpha^{2}}\frac{e^{2\alpha \xi(t)}}{\left(\int \limits _{0}^{t} \,e^{\alpha\xi(\tau)}d\tau\right)^{2} }\,.
\end{equation}
Note that the explicit dependence  on the parameter $\lambda$
is missing in (\ref{Eq:V}).

As $V(\xi(t))$ does not depend on the moments of time exceeding $t\,,$
the Wiener integration in the r.h.s. of the equation (\ref{Eq:Equality}) over the space of functions
$\xi \in C([0,\,+\infty)) $ is reduced to the Wiener integration over the space $ C([0,\,t])\,. $

Therefore, $\langle V(t)\rangle$ has the obvious meaning. It is nothing but the mean value of the function of Wiener process at the moment of time $t\,.$

Thus, we have
\begin{equation}
   \label{Eq:V1}
\langle V(t)\rangle=-\frac{1}{2} I_{1}(t)+\frac{1}{2\alpha^{2}}I_{2}(t)\,,
\end{equation}
where
\begin{equation}
   \label{Eq:I1}
I_{1}(t)=\int\limits _{\small{\begin {matrix} C([0,t]) \\ \xi(0)=0 \end {matrix}}} \frac{e^{\alpha\xi(t)}}{\int \limits _{0}^{t} \,e^{\alpha\xi(\tau)}d\tau } \, w_{1}(d\xi)\,,
\end{equation}
and
\begin{equation}
   \label{Eq:I2}
I_{2}(t)=\int\limits _{\small{\begin {matrix} C([0,t]) \\ \xi(0)=0 \end {matrix}}} \frac{e^{2\alpha\xi(t)}}{\left(\int \limits _{0}^{t} \,e^{\alpha\xi(\tau)}d\tau\right)^{2} } \, w_{1}(d\xi)\,.
\end{equation}

The substitution
$$
\eta(\tau)=\alpha\xi(t\tau),\ \ \ \tau\in[0,1]\,,
$$
and the identity
$$
\int\,F[\eta(1), \eta]\,w_{\sigma}(d\eta)
$$
$$
=\int\left(\int\limits_{-\infty}^{+\infty}\,\delta(\eta(1)-x)\, F[x,\eta]\, dx\right) \, w_{\sigma}(d\eta)
$$
\begin{equation}
   \label{Eq:Identity}
=\int\limits_{-\infty}^{+\infty}\,\left(\int\limits _{\small{ \eta(1)=x }} F[x,\eta]\, w_{\sigma}(d\eta)\right)\,dx
\end{equation}
make it possible to rewrite the integrals $I_{1}\,,$ $I_{2}$ in the form
\begin{equation}
   \label{Eq:I10}
I_{1}(t)=\int\limits_{-\infty}^{+\infty}\,\frac{e^{x}}{t}\left(\int\limits _{\small{\begin {matrix} C([0,1]) \\ \eta(0)=0, \\ \eta(1)=x \end {matrix}}}\frac{1}{\int \limits _{0}^{1} \,e^{\eta(\tau)}d\tau } \, w_{\sigma}(d\eta)\right)\,dx\,,
\end{equation}
\begin{equation}
   \label{Eq:I20}
I_{2}(t)=\int\limits_{-\infty}^{+\infty}\,\frac{e^{2x}}{t^{2}}\left(\int\limits _{\small {\begin {matrix} C([0,1]) \\ \eta(0)=0, \\ \eta(1)=x \end {matrix}}}\frac{1}{\left(\int \limits _{0}^{1} \,e^{\eta(\tau)}d\tau \right)^{2}} \, w_{\sigma}(d\eta)\right)\,dx\,,
\end{equation}
where
$
\sigma=|\alpha|\,\sqrt{t} \,.
$

The above Wiener integrals (\ref{Eq:I10}) and (\ref{Eq:I20}) are the first two terms in the expansion of the generating integral
$$
\int\limits _{\small{\begin {matrix} C([0,1]) \\ \xi(1)=x \end {matrix}}}\exp\left\{ - \frac{2\beta}{\sigma^{2}}\,\left[1-\frac{e^{x}}{\beta+1} \right]\,\frac{1}{\int \limits _{0}^{1} \,e^{\xi(\tau)}d\tau } \right\}\, w_{\sigma}(d\xi)
$$
\begin{equation}
   \label{Eq:Formula}
=\frac{1}{\sqrt{2\pi} \sigma}\, \exp \left\{ - \frac{\left[x-2\ln (\beta+1) \right]^{2}}{2\sigma ^{2}}   \right\} \,.
\end{equation}
A proof of the equation (\ref{Eq:Formula}) (too lengthy one for the letter) can be obtained using the technique developed in \cite{BSh}, \cite{Sh}.

Thus, the results are
\begin{equation}
   \label{Eq:I12}
I_{1}(t)=\frac{|\alpha|}{\sqrt{2\pi\,t}}\int\limits_{0}^{+\infty}\,x\,\coth    \left( |\alpha|\sqrt{t} \frac{x}{2}\right) \exp ^{-\frac{x^{2}}{2}}\,dx\,,
\end{equation}
and
$$
I_{2}(t)=\frac{\alpha^{2}}{\sqrt{2\pi}\ t}\int\limits_{-\infty}^{+\infty}\,\frac{1}{\left(1-e^{-|\alpha|\sqrt{t} x} \right)^{2}}
$$
\begin{equation}
   \label{Eq:I22}
\times\left\{x^{2}-1+|\alpha|\sqrt{t}\frac{x}{2} \,\coth    \left( |\alpha|\sqrt{t} \frac{x}{2}\right) \right\}\exp ^{-\frac{x^{2}}{2}}\,dx\,.
\end{equation}

\section{4 The asymptotic behaviour of the model}
\label{sec:R}

We postpone the detailed analysis of numerical solutions of the Riccati equation (\ref{Eq:Riccati}) for different values of the parameter $\alpha $ until a forthcoming paper. And here, we study the asymptotic behaviour of the solution at $t\rightarrow +0 $ and $t\rightarrow +\infty\,.$

However, first we make a general comment. A graph of the averaged potential has a typical depression (FIG.1).
\begin{figure}
\includegraphics [height=0.3\textwidth]{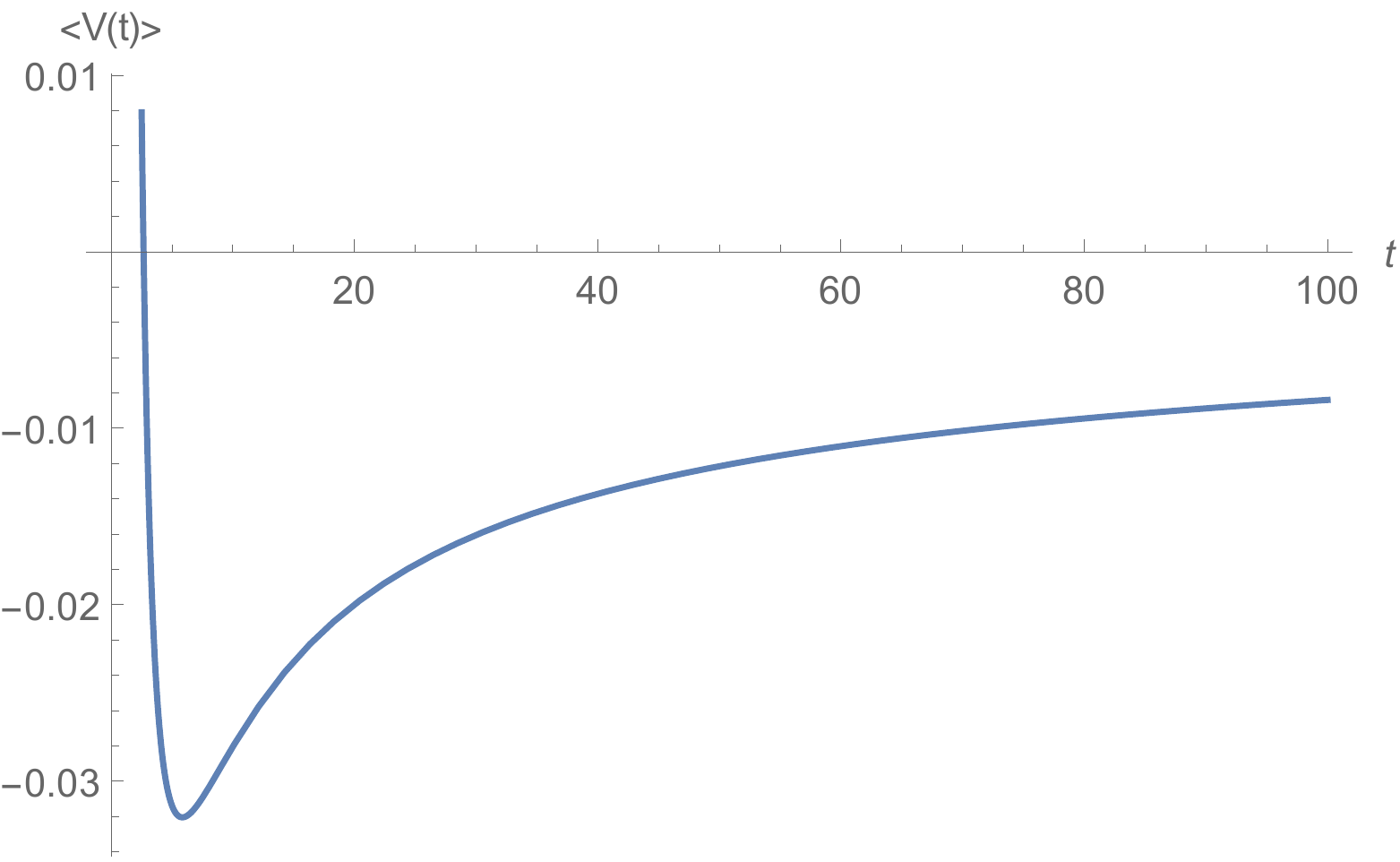}
\caption{The form of the averaged potential $\langle V(t)\rangle $  $(|\alpha |=\sqrt{\frac{2}{3}})\,.$}
\end{figure}
So, there are solutions of the Riccati equation with
$$
\dot{H}(t)>0\,, \ \ \ddot{a}(t)>\frac{\dot{a}^{2}(t)}{a(t)}
$$
in this region.

From (\ref{Eq:I12}), (\ref{Eq:I22}) and (\ref{Eq:V1}), we deduce that
$$
\langle V(t)\rangle\simeq \frac{1}{2\alpha^{2}}\frac{1}{t^{2}}
$$
at $t\rightarrow+0\,.$

The asymptotic Riccati equation looks like
\begin{equation}
\label{Eq:Riccati1}
\dot{H}_{0}(t)+3H_{0}^{2}(t)=-\frac{1}{2\alpha^{2}}\frac{1}{t^{2}}\,.
\end{equation}

In particular, for $\alpha^{2}\geq 6\,,$ it has solutions of the form
\begin{equation}
\label{Eq:Sol1}
H_{0}(t)=\frac{q(\alpha )}{t}\,,\ \ \ \ q (\alpha )=\frac{1}{6}\pm\sqrt{1-\frac{6}{\alpha^{2}}}\,.
\end{equation}

So, at small $t\,,$ the scale factor behaves as
\begin{equation}
\label{Eq:Beh1}
a(t)\sim t^{q(\alpha )}\,.
\end{equation}

Note that due to the nonlinearity of the Riccati equation, there are also other asymptotic solutions besides (\ref{Eq:Sol1}).

 At $t\rightarrow+\infty \,, $ the asymptotic behavior of our model with nonzero $\Lambda $ is the same as the one of the model without the scalar field.
Therefore, we suppose that $\Lambda =0\,.$

In this case,
$$
\langle V(t)\rangle\simeq -\frac{|\alpha |}{4\sqrt{2\pi}}\,t^{-\frac{1}{2}}
$$
at $t\rightarrow+\infty\,.$

The asymptotic Riccati equation has the form
\begin{equation}
\label{Eq:E-Riccati2}
\dot{H}_{\infty}(t)+3H_{\infty}^{2}(t)=\frac{|\alpha |}{4\sqrt{2\pi}}\,t^{-\frac{1}{2}}\,.
\end{equation}

There are two different asymptotic solutions:
\begin{equation}
\label{Eq:Sol2}
H_{\infty}(t)=\pm\frac{1}{2}\sqrt{\frac{|\alpha |}{3\sqrt{2\pi}}}\,t^{-\frac{1}{4}}\,.
\end{equation}
( In this case, the term $\dot{H}_{\infty}(t) \sim t^{-\frac{5}{4}}$ in (\ref{Eq:E-Riccati2}) is negligible.)

Thus, for the $"+"$ sign in the r.h.s. of (\ref{Eq:Sol2}), the scale factor is growing asymptotically as
\begin{equation}
\label{Eq:FutureSF1}
a_{\infty}(t)=C\,\exp\left\{ \frac{2}{3}\sqrt{\frac{|\alpha |}{3\sqrt{2\pi}}}\,t^{\frac{3}{4}}\right\}\,.
\end{equation}
For the $"-" $ sign, it is vanishing
\begin{equation}
\label{Eq:FutureSF2}
a_{\infty}(t)=C\,\exp\left\{- \frac{2}{3}\sqrt{\frac{|\alpha |}{3\sqrt{2\pi}}}\,t^{\frac{3}{4}}\right\}\,.
\end{equation}

\section{5 Conclusion}
\label{sec:concl}

In spite of the simplicity of the proposed model, it leads to rather nontrivial solutions.
The main advantage of our approach consists in that we are able to find the explicit form of the function $\langle V(t)\rangle $ on the axis $0<t<+\infty$ and use it in the equation for the scale factor.

This paper is a first attempt to take into account the quantum effects of the singular scalar field  on the scale factor as a whole.
The point is that the traditional way to study the quantum effects perturbatively is nonapplicable in the presence of singularities. Even the quasi-classical limit of the theory with a singular field is different from that given by the classical action \cite{BSh}, \cite{QR}.

Of course, we are very far from the complete quantum description of gravity interacting with matter fields. The state of the art reminds the early days of quantum theory when the electromagnetic field was treated as a quantum one but the matter was considered classicaly.

\section{Acknowledgements}
\label{sec:acknowl}
The authors are grateful to V.A. Rubakov for stimulating discussion and enlightening comments.

\end{document}